\documentclass[aps,prl,reprint,showpacs,showkeys,superscriptaddress,floatfix,twocolumn]{revtex4-2}
\usepackage{multirow}
\usepackage{graphicx}
\usepackage{bm}
\usepackage{dcolumn}
\usepackage[colorlinks=true,linkcolor=blue,urlcolor=blue,citecolor=blue]{hyperref}
\usepackage{natbib}
\usepackage{amsmath}
\usepackage{times}
\usepackage{xcolor}
\usepackage{soul}
\usepackage{chemformula}
\usepackage[normalem]{ulem}

\begin{document}

\title{{Emergent Hidden Multipolar State in the Triangular Lattice Magnet \ch{TmZn2GaO5}}}

\author{Matthew Ennis}\thanks{equal contribution}\affiliation{Department of Physics, Duke University, Durham, NC, USA}
\author{Rabindranath Bag}\thanks{equal contribution}\affiliation{Department of Physics, Duke University, Durham, NC, USA}
\author{Tessa Cookmeyer}\thanks{equal contribution}\affiliation{Kavli Institute for Theoretical Physics, University of California, Santa Barbara, CA, USA}
\author{Matthew B. Stone}\affiliation{Neutron Scattering Division, Oak Ridge National Laboratory, Oak Ridge, TN, USA}
\author{Alexander I. Kolesnikov}\affiliation{Neutron Scattering Division, Oak Ridge National Laboratory, Oak Ridge, TN, USA}
\author{Tao Hong}\affiliation{Neutron Scattering Division, Oak Ridge National Laboratory, Oak Ridge, TN, USA}
\author{Leon Balents}\affiliation{Kavli Institute for Theoretical Physics, University of California, Santa Barbara, CA, USA}
\author{Sara Haravifard}
\email[email:]{sara.haravifard@duke.edu} \affiliation{Department of Physics, Duke University, Durham, NC, USA} \affiliation{Department of Mechanical Engineering and Materials Science, Duke University, Durham, NC, USA}
\affiliation{Department of Electrical and Computer Engineering, Duke University, Durham, NC, USA}


\begin{abstract}
\ch{TmZn2GaO5} is a newly synthesized triangular lattice magnet that exhibits a unique quantum phase characterized by strong Ising anisotropy, a pseudo-doublet crystal electric field ground state, and a low-energy gapped excitation at the K point. Unlike its well-known counterparts, \ch{TmMgGaO4} and \ch{YbMgGaO4}, this material crystallizes in a distinct hexagonal structure, leading to a cleaner platform for investigating frustrated magnetism. Magnetic susceptibility, heat capacity, and inelastic neutron scattering measurements confirm the absence of long-range magnetic order down to 50 mK, placing \ch{TmZn2GaO5} in a distinct region of the transverse-field Ising model phase diagram. Theoretical calculations based on spin-wave theory and mean-field modeling reproduce key experimental observations, reinforcing the material's placement in a quantum disordered/multipolar state. These results highlight its potential for exploring quantum disordered states, anisotropic excitations, and exotic quantum phases in frustrated spin systems.
\end{abstract}
	
\maketitle

\textit{Introduction}--Over the last few decades, materials exhibiting magnetic frustration have attracted considerable attention for their potential to host exotic states, such as spin ices \cite{Bramwell_2001, Ross_2011, Gingras_2014} and quantum spin liquids (QSLs) \cite{Savary_2016, Zhou_2017, Broholm_2020}. Among these, materials with antiferromagnetic spins arranged on a triangular lattice have been particularly attractive due to their intrinsic geometric frustration \cite{Leung_1993, Ramirez_1994, Collins_1997, Starykh_2015}. Since the early 2010s, \ch{YbMgGaO4}, a rare-earth system with triangular lattice geometry, has been extensively studied as a potential host of a QSL state. \ch{YbMgGaO4} shows behavior consistent with a U(1) spin liquid ground state \cite{Li_2015, li2015rare, Li_2016_YMGO, Paddison_2017, Li_2017,Shen_2018, Li_2019_YMGO, Majumder_2020, Rao_2021}, however, presence of significant site-mixing disorder between the non-magnetic Mg and Ga atoms complicates the interpretation of the experimental results \cite{Xu_2016, Li2017crystalline, Bachus_2020}. This has led to several other explanations of the measurements, and the true ground state remains uncertain \cite{Zhu_2017, Parker_2018, Kimchi_2018}. Similarly, the isostructural compound \ch{YbZnGaO4} has also been investigated, with some studies reporting signatures of a spin liquid, while others suggest that the site mixing drives it into a spin glass ground state \cite{Ma_PRL_2018, Ma_2021, Steinhardt_2021, Pratt_2022}.

Recently, the sister compound \ch{TmMgGaO4} has been synthesized and studied. Initial magnetic measurements on single crystal samples revealed significant anisotropy, attributed to crystal electric field (CEF) interactions \cite{Cevallos_2018}. Although Tm$^{3+}$ is not a Kramers ion, the CEF ground state was indentified as a pseudo-doublet formed by two closely spaced singlets \cite{Li_PRX_2020, Dun_PRB_2021}. This CEF ground state can be modeled with a transverse-field term in the Hamiltonian, which contributes to the large anisotropy \cite{Shen2019, Li_PRX_2020, Liu_2020, Li_NatComm_2020, Dun_PRB_2021}. At low temperatures, \ch{TmMgGaO4} orders into a 3-sublattice phase \cite{Li_NatComm_2020}, confirmed by neutron scattering experiments showing magnetic Bragg peaks at the K points \cite{Shen2019, Dun_PRB_2021}. 

At finite temperatures, a possible transition to a Berezinskii-Kosterletz-Thouless (BKT) phase has been proposed \cite{Berezinskii_1971, Berezinskii_1972, Kosterlitz_1973}, and recent experiments support the formation of this phase in \ch{TmMgGaO4} \cite{Li_NatComm_2020, Hu_2020, Dun_PRB_2021}. Unlike \ch{YbMgGaO4}, where site disorder plays a significant role in the underlying physics, it appears that disorder does not have as strong of an effect on the ground state of \ch{TmMgGaO4} \cite{Hu_2020, Dun_PRB_2021}. However, this conclusion is not universal, as a recent report argues that the ground state is a U(1) gauge glass due to structural disorder in \ch{TmMgGaO4} \cite{Huang_2024}.

The study of these materials highlights the importance of a disorder-free materials in the search for these exotic phases. Motivated by this, we have begun synthesis and characterization of the related compound \ch{TmZn2GaO5}. Unlike \ch{TmMgGaO4}, \ch{TmZn2GaO5} crystallizes in a different space group and minimizes site mixing between the nonmagnetic Zn and Ga atoms due to their distinct Wyckoff positions. We successfully grew single crystal samples of \ch{TmZn2GaO5} using the floating zone method and conducted magnetic, heat capacity, and inelastic neutron scattering measurements to explore its ground state physics. Recently, we reported a similar study on the sister compound \ch{YbZn2GaO5}, where we identified evidence of a potential U(1) Dirac quantum spin liquid state \cite{Yb1215}. 

In this letter, we present our experimental results and theoretical insights on \ch{TmZn2GaO5}. Magnetic susceptibility and heat capacity measurements show no sign of magnetic ordering. Inelastic neutron scattering reveals a low-energy gapped excitation originating from the K point. By fitting the data to the transverse-field Ising model, we demonstrate that \ch{TmZn2GaO5} occupies a distinct region of the phase diagram, and does not exhibit the 3-sublattice ordered phase previously observed in \ch{TmMgGaO4}, instead forming a ``hidden'' multipolar state which is not observable through neutron scattering \cite{Santini_2009, Mydosh_2011}.

\begin{figure}
\centering
\includegraphics[width=9 cm]{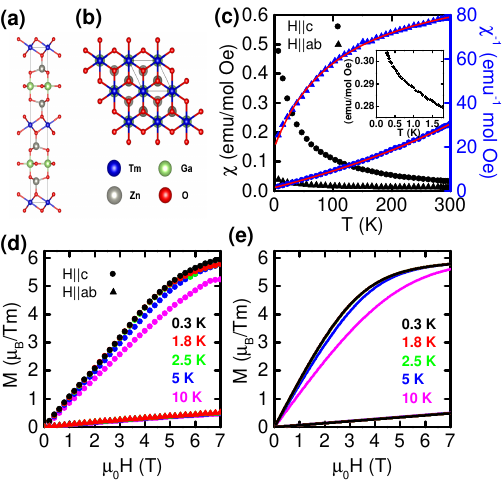}
\caption{(a) Crystal structure of \ch{TmZn2GaO5} viewed along $a$-axis: showing well separated magnetic Tm-O layers along $c$-axis (b) Crystal structure viewed along the $c$-axis, showing the triangular arrangement of Tm atoms in $ab$-plane (c) Temperature dependent magnetic susceptibility (black) and inverse susceptibility (blue) for single crystals with the field (H = 0.01 T) applied in-plane (circles, H $\parallel$ $ab$) and out-of-plane (triangles, H $\parallel$ $c$). The red lines represent fits to the Curie-Wess law (see text). The inset shows the helium-3 magnetic susceptibility data collected on a powder sample of \ch{TmZn2GaO5}, indicating no long range magnetic ordering down to 0.3 K (d) Isothermal magnetization with field applied in-plane (triangles) and out-of-plane (circles) (e) Mean-field simulation of isothermal magnetization at various temperatures along the $c$-axis (upper) and in the $ab$-plane (lower).}
\label{fig:mag}
\end{figure}

Powder samples of \ch{TmZn2GaO5} were synthesized using solid-state reactions route. High purity precursors of \ch{Tm2O3}, \ch{ZnO}, and \ch{Ga2O3} were mixed in a 1:1.05:1 ratio by weight to account for ZnO evaporation and to prevent unreacted \ch{Tm2O3} in the final product. This mixture was pressed into pellets and sintered in air at 1100 $^\circ$C for 12 hours, followed by 1350 $^\circ$C for 60 hours. The phase and purity of the sintered powders were confirmed using powder X-ray diffraction (See supplementary information for details \cite{SM}). Once the pure powder samples were obtained, they were pressed into cylindrical feed and seed rods under hydrostatic pressure of 700 bar and grown into single crystals using a four-mirror optical floating zone furnace equipped with xenon arc lamps. The single crystal grains were oriented and extracted using the Laue X-ray diffraction for subsequent magnetic, heat capacity and neutron measurements. We used VESTA \cite{VESTA} to visualize and plot the crystal structure as shown in Fig. \ref{fig:mag}(a,b). Magnetic measurements were performed down to 0.3 K with fields up to 7 T on single crystal samples using a SQUID magnetometer with a helium-3 attachment. Heat capacity measurements were performed down to 0.06 K on a single crystal of \ch{TmZn2GaO5} using a Quantum Design Physical Property Measurement System (PPMS) with a dilution refrigerator attachment. Inelastic neutron scattering  measurements were carried out on the SEQUOIA time-of-flight spectrometer \cite{SEQUOIA} at Oak Ridge National Laboratory to study the crystal electric field (CEF) spectrum. The neutron scattering data were analyzed using DAVE \cite{DAVE}, and the Python package PyCrystalField \cite{PyCrystalField}. The single crystal inelastic neutron scattering measurements were performed on CTAX at Oak Ridge National Laboratory to investigate the low-energy dispersion observed in the SEQUOIA data. The final neutron energy of CTAX was fixed at 3.5 meV and a cooled Be filter was placed after the sample to eliminate higher-order beam contamination.

\ch{TmZn2GaO5} crystallizes in the hexagonal space group $P6_3/mmc$ (no. 194). This is in contrast to the previously reported compound \ch{TmMgGaO4}, which forms the trigonal space group $R\Bar{3}m$ \cite{Cevallos_2018, Ma_PRL_2018}. The crystal structure of \ch{TmZn2GaO5} is shown in Fig. \ref{fig:mag}(a), and the triangular arrangement of Tm atoms is shown in \ref{fig:mag}(b). Unlike \ch{TmMgGaO4}, the presence of an additional \ch{ZnO} in the formula unit of \ch{TmZn2GaO5} increases the separation between Tm layers. The inclusion of the \ch{ZnO} also alters the space group, placing Zn and Ga atoms at different Wyckoff positions (4f and 2b, respectively). This change in structure significantly reduces the site mixing that affects \ch{TmMgGaO4} and \ch{YbMgGaO4}. The Rietveld refinement of the powder X-ray diffraction data of \ch{TmZn2GaO5} confirmed the $P6_3/mmc$ phase (see, Supplemental Fig. 1).

Temperature dependent magnetic susceptibility and inverse magnetic susceptibility along different crystallographic directions down to 1.8 K are shown in Fig. \ref{fig:mag}(c). The inset shows susceptibility of a powder sample down to 300 mK, and we do not see any signatures of long range magnetic ordering. Significant easy-axis anisotropy is observed along the crystallographic $c$-direction. The magnetic susceptibility data were fitted to the Curie-Weiss law: $\chi = \frac{C}{T-\theta_{W}} + \chi_0 $; where $\chi_0$ is a temperature independent susceptibility term, $C$ is the Curie constant, and $\theta_{W}$ is the Weiss temperature. By adding the temperature-independent term we were able to fit the data over the entire temperature range. For both orientations, the fits reveal large negative Weiss temperatures ($\theta^c_{W} = -24.40$ K and $\theta^{ab}_{W} = -22.37$ K ), indicating strong antiferromagnetic interactions. With the field applied along the $c$-axis, the temperature-independent susceptibility is $\chi_0^c = -0.0147 \ \mathrm{emu/mol\text{-}Tm \ Oe}$, while for the in-plane field, it is slightly smaller and positive value, $\chi_0^{ab} = 0.0098 \ \mathrm{emu/mol\text{-}Tm \ Oe}$. The core diamagnetism for \ch{TmZn2GaO5} can be computed from reference tables \cite{Bain_Berry_2008} which gives a value of $\chi_D = -116\times 10^{-6}$ emu/mol Oe, far too small to account for the large temperature-independent susceptibility observed along the $c$-axis. Interestingly, a similar trend is observed on the previously studied compound \ch{TmMgGaO4} \cite{Cevallos_2018, Shen2019, Li_NatComm_2020}, but its origin remains unexplained. 

\begin{figure*}[!]\centering\includegraphics[width=18 cm]{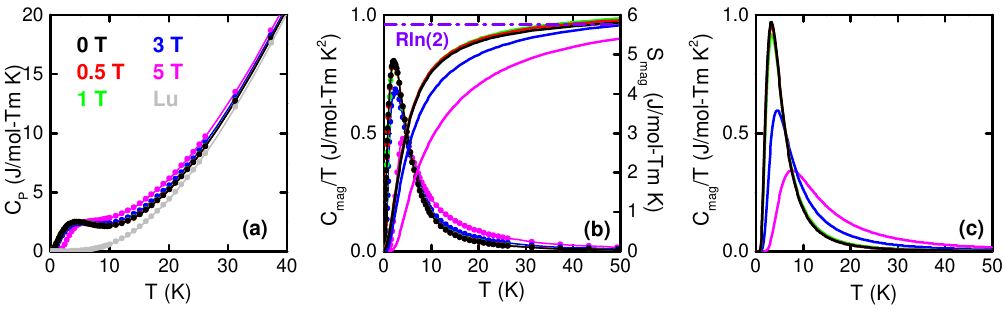}
	\caption{(a) Heat capacity of a single crystal sample of \ch{TmZn2GaO5} with field applied along the $c$-axis, compared with the nonmagnetic \ch{LuZn2GaO5} (grey points, under zero field) for reference (b) Magnetic heat capacity (points, left axis) and calculated magnetic entropy (lines, right axis). The dashed magneta line indicates the position of $\mathrm{R} \ln(2)$ on the right axis (c) Mean-field calculation of magnetic heat capacity. }
	\label{fig:heatcap}
\end{figure*}  

To further understand the magnetic behavior of \ch{TmZn2GaO5}, isothermal magnetization measurements were performed on single crystal samples at various temperatures up to 7 T. The magnetization data along different crystallographic directions are shown in Fig. \ref{fig:mag}(d). Consistent with the susceptibility data, significant anisotropy is observed, with the magnetization along $c$-direction being more than 10 times larger than in the plane. The in-plane magnetization follows a Brillouin function profile, however the slopes of the magnetization curves along the $c$-direction exhibit some nontrivial features. Notably, the derivative $dM/dH$ displays a minimum at 1.5 T and a maximum at 3 T, suggesting complex magnetic interactions along this direction (see Supplementary Fig. 2(b)). The magnetization in both directions is relatively temperature insensitive. As shown in Fig. \ref{fig:mag}(d), at 1.8 K and 2.5 K, the magnetization curves are nearly identical, and even at 5 K, they remain very similar. A significant decrease in magnetization values is only observed at 10 K. We also simulated the magnetization using the mean-field model described below for the same temperatures that were measured. The simulation results, shown in Fig. \ref{fig:mag}(e), quantitatively agree with the experimental data, capturing the overall shape and temperature dependence observed in our SQUID measurements. This alignment supports the applicability of the mean-field model to describe the magnetic properties of \ch{TmZn2GaO5}. 

To gain further insight into the system, low temperature heat capacity measurements were performed on a single crystal sample of \ch{TmZn2GaO5} along the crystallographic $c$-direction. No long range ordering was observed down to 0.06 K. The heat capacity data up to 50 K is shown in Fig. \ref{fig:heatcap}(a), with the nonmagnetic analogue \ch{LuZn2GaO5} included for reference. A broad peak is observed around 4.5 K. This peak superficially resembles a two-level Schottky peak; however, the peak is too broad and in particular the high-temperature tail decays too slowly to be adequately described by the Schottky formula. The peak is also relatively resilient to applied magentic field, exhibiting behavior similar to the magnetization with respect to temperature. Specifically, the peak's width and position show minimal changes up to fields of 1 T. At 3 T, the peak broadens slightly and shifts to higher temperatures, with this effect becoming more pronounced at 5 T. The magnetic heat capacity per temperature, obtained by subtracting the lattice contribution estimated from the heat capacity of the isostructural nonmagnetic \ch{LuZn2GaO5}, and the corresponding magnetic entropy are shown in Fig. \ref{fig:heatcap}(b). Below 100 mK we also observed a small nuclear Schotty peak which we have subtracted off. We find that the magnetic entropy saturates to $\mathrm{R} \ln(2)$ by 50 K, indicating an effective spin-$\frac{1}{2}$ ground state. The magnetic heat capacity was modeled using Eq. \ref{eq:model}, as shown in Fig. \ref{fig:heatcap}(c). The calculated mean-field results align well with the experimental data, capturing the trends where the heat capacity remains largely unchanged up to 1 T and deviates significantly only at fields of 3 T and above.

\begin{figure*}[htbp]\centering\includegraphics[width=15cm]{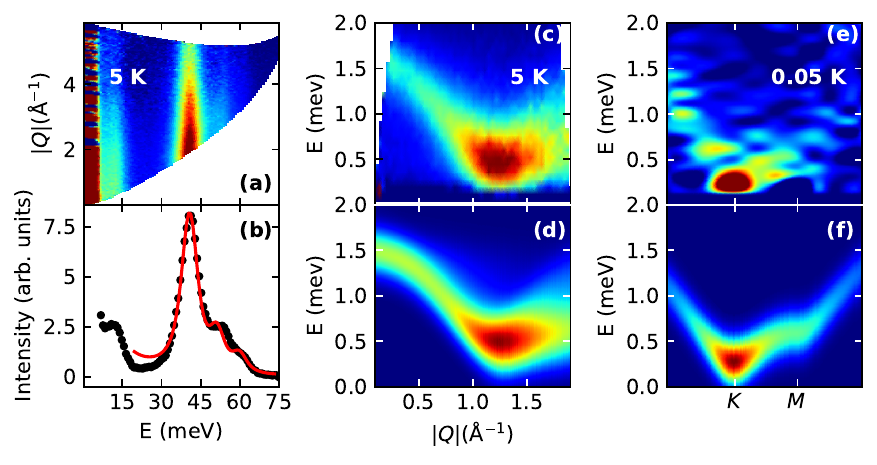}
	\caption{(a) Inelastic neutron scattering data on a powder sample of \ch{TmZn2GaO5} using \ch{LuZn2GaO5} for background subtraction (b) Integrated intensity (black points) and the fitted CEF model (red solid line). (c) Low-energy inelastic neutron powder data from SEQUOIA (d) spin-wave theory calculation of powder data shown in c. (e) Inelastic neutron scattering data on single crystals of \ch{TmZn2GaO5} from CTAX (f) spin-wave theory calculation of single crystal data.}
	\label{fig:neutron}
\end{figure*}

To further investigate the ground state of \ch{TmZn2GaO5}, we performed the inelastic neutron scattering (INS) measurements on a powder sample using the SEQUOIA spectrometer \cite{SEQUOIA}. We used incident neutron energies 5 meV, 8 meV, 80 meV, and 150 meV, and conducted measurements at both 5 K and 100 K for each energy. To account for the phonon contribution, we also measured the nonmagnetic \ch{LuZn2GaO5} under the same conditions. Fig. \ref{fig:neutron}(a) shows the 5 K INS data with 80 meV incident energy, while the data for the other temperatures and incident energies are provided in Supplemental Fig. 3. We observed a large band with magnetic form factor around 40 meV, along with several weaker bands at higher energies. Additionally, we observed a fainter band around 13 meV, however, we believe that this feature is not due to CEF and is likely to be a sample-dependent feature, as it was not observed on a repeat measurement with a different powder sample. 

To determine the CEF parameters, we integrated the neutron scattering data between 1.5 \AA$^{-1}$ and 4.5 \AA$^{-1}$ to obtain the intensity as a function of energy transfer, as shown in Fig. \ref{fig:neutron}(b). We analyzed the data using the Stevens operator formalism described by Hutchings \cite{Hutchings1964}. In \ch{TmZn2GaO5}, the Tm atoms belong to the point group $D_{3d}$, which allows only six nonzero CEF parameters \cite{Walter_1984}. Thus, we fit the CEF spectrum using the following Hamiltonian:

\begin{equation}
    \begin{aligned}
        \mathcal{H}_\mathrm{CEF} = &B_2^0\mathcal{O}_2^0 + B_4^0\mathcal{O}_4^0 + B_4^3\mathcal{O}_4^3 \\ &+ B_6^0\mathcal{O}_6^0 + B_6^3\mathcal{O}_6^3 + B_6^6\mathcal{O}_6^6
    \end{aligned}
    \label{eq:CEFham}
\end{equation}

\noindent Where $\mathcal{O}_m^n$ are the Stevens operators \cite{Stevens_1952} and $B_m^n$ are the corresponding CEF parameters. Tm$^{3+}$ has spin-1, so Kramer's theorem does not apply and the CEF levels can be a mix of singlets and doublets, which complicates identification of the energy levels. Furthermore, 10 energy levels are expected but only 3 distinct peaks are observed (at 40 meV, 55 meV, and 60 meV). We have measured with up to 150 meV neutrons and do not observe any clear excitations between 90 meV and 150 meV. Additionally, no CEF excitations are observed below 30 meV. Our heat capacity measurements show that the magnetic entropy saturates to Rln(2), indicating that the CEF ground state must be a doublet or two close-lying singlets. Guided by these observations, we fit the data to Eq. \ref{eq:CEFham} using the Python package PyCrystalField \cite{PyCrystalField}. Starting from a point charge estimation of the parameters, the 5 K data with 80 meV and 150 meV incident energy were fit simultaneously. The results of the fit are also shown in Fig. \ref{fig:neutron}(b). The best fit was obtained with a two-singlet ground state, with the higher-energy singlet at  0.15 meV, leading to a pseudo-doublet ground state. The ground state is composed mostly of the $|\pm 6\rangle$ and $|\pm 3\rangle$ basis vectors, which explains the large anisotropy observed in the magnetic measurements. The CEF parameters and energy levels are provided in Supplemental Tables I and II. We found that the CEF excitations were significantly broader than the instrumental resolution, similar to previous reports on \ch{TmMgGaO4} and \ch{YbMgGaO4} \cite{Dun_PRB_2021, Paddison_2017, Li_2017}.

While measuring with low incident energy neutrons, we observed what appears to be a dispersive excitation below 2 meV. The 5 K measurement with 8 meV neutron incident energy is shown in Fig. \ref{fig:neutron}(c). This excitation appears to originate from the K point and shows a slight gap of approximately 0.1 meV. However, this value is at the limits of the energy resolution for this experiment, making precise determination challenging. To further investigate this excitation, we conducted inelastic neutron scattering measurements on four coaligned single crystal samples of \ch{TmZn2GaO5} in the HK0 scattering plane using CTAX. Measurements were performed at 0.05 K, with energy transfer up to 2 meV and magnetic fields up to 8 T at points within the first Brillouin zone along the path $\Gamma-M-K-\Gamma$. Due to the small dimensions of the \ch{TmZn2GaO5} unit cell, we were unable to access points outside the first Brillouin zone. We found that the excitation does not get lifted on applying field (as would be expected for a simple spin-wave excitation, for example) but instead becomes suppressed. By 8 T, the excitation had completely dissipated, so we used the 0.05 K measurement at 8 T for background subtraction in lieu of a high-temperature measurement. The zero field measurement at 0.05 K is shown in Fig. \ref{fig:neutron}(e).

Our single crystal data reveals that the excitation emerges from the K point. The excitation appears to be slightly gapped, with the most intense part of the excitation lying around 0.3 meV. Along the path from the K point to the M point, the band rises slightly and plateaus around 0.6 meV. From both the K and M points to the central $\Gamma$ point the band rises sharply to about 1.5 meV. This dispersion behavior is qualitatively similar to what has been observed in \ch{TmMgGaO4} \cite{Shen2019}, with the important distinction that the excitation in \ch{TmZn2GaO5} is gapped at the K points while the one in \ch{TmMgGaO4} is gapless. To confirm this observation, we also performed elastic rocking curve measurements on CTAX at the K and M points. We did not observe any peaks or other features at either the K or M points, as shown in Supplemental Figure 4. This is in contrast to \ch{TmMgGaO4}, where an elastic peak is observed at the K point \cite{Shen2019, Dun_PRB_2021}. \ch{TmMgGaO4} also exhibits an elastic peak at the $\Gamma$ point. Although we could not directly measure the $\Gamma$ point in \ch{TmZn2GaO5}, our model predicts that such a peak should also be present in this material.

In order to understand the previous measurements, we study the model Hamiltonian introduced in Ref.~\cite{Shen2019}. We describe the pseudo-doublet ground state as an effective spin-$1/2$ moment with small splitting set by an intrinsic transverse field $\Delta$. Since the ground state is mostly composed of the $\pm |6\rangle$ states, the natural spin-spin coupling is of Ising-type, rather than Heisenberg, and we assume that the couplings are predominantly antiferromagnetic. Importantly, neutron scattering only directly couples to the $S^z$ component of the spins \cite{Shen2019,SM}. 

In total, we have the model Hamiltonian -- an extended quantum transverse field Ising model -- for spin-$1/2$s on a triangular lattice:

\begin{equation}
    \mathcal{H} = \sum_{\langle ij \rangle} J_1S^z_iS^z_j + \sum_{\langle \langle ij \rangle \rangle} J_2S^z_iS^z_j + \sum_i \left(\Delta S^y_i - h_zS^z_i\right)
    \label{eq:model}
\end{equation} 

\noindent where $h^z = g_\text{eff}\mu_B H^z$ for the external field $H^z$, Bohr magneton $\mu_B$, and effective $g$-factor $g_\text{eff}$. The magnetic field in the $H^x$ and $H^y$ does not directly couple to $S^x$ and $S^y$ as they are even under time-reversal \cite{Shen2019,SM}.

At the classical level, with no magnetic field, this model has three phases (Fig.~\ref{fig:phasediagram}) (see the Supplemental Material \cite{SM}).  When $J_2\approx 0$ and $J_1/\Delta \gtrsim 0.6$, there is a three-sublattice order in the $y$-$z$ plane with all spins predominantly pointing in the $y$ direction but where one spin is canted above the $x$-$y$ plane and another canted below the $x$-$y$ plane. When $J_2$ is large enough, a two-sublattice ``stripe'' order is preferred, where one spin cants above the $x$-$y$ plane and the other below.  Finally, when $J_1,J_2 \ll \Delta$, the spins are polarized in the $\hat y$ direction: this is the ``quantum disordered'' phase of the transverse field Ising model, and reflects a multipolar state in physical terms. 

In Ref.~\cite{Shen2019}, the parameters are such that the three-sublattice order is preferred, which then predicts an elastic peak at the $K$ point. For \ch{TmZn2GaO5}, no such peak is observed implying that the material must be in the polarized phase. We carry out a spin-wave calculation to find the dispersion relation for such a state \cite{SM}. We estimate the parameters by matching the single-crystal data and the spin-wave dispersion at the $\Gamma$, $K$, and $M$ point. We find $(J_1,J_2,\Delta)=(0.53 \text{ meV},-0.01 \text{ meV},0.91 \text{ meV})$. These parameters place \ch{TmZn2GaO5} close to the phase boundary between the quantum disordered/multipolar state and the three-sublattice ordered phase. Using these parameters, we calculate the powder [Fig.~\ref{fig:neutron}(d)] and single-crystal [Fig.~\ref{fig:neutron}(e)] INS spectra using the \texttt{spinw} package \cite{toth2015linear}, which agrees quantitatively with the experiment. 

We can further evaluate the magnetization and specific heat within a mean-field calculation. We determine that $g_\text{eff}\approx 12$, within our model, in order to reproduce the magnitude of the magnetization data, after which there are no more free parameters. We see in Fig.~\ref{fig:mag}(e) and Fig.~\ref{fig:heatcap}(c) that the mean-field calculation quantitatively captures both the magnetization and specific heat data. The specific heat peak is due to a Schottky anomaly between the quasi-doublet states and not due to a phase transition. 

The in-plane magnetization data can be reproduced within a mean-field calculation considering the full $J=6$ moments for the Tm$^{3+}$. In the SM \cite{SM}, we consider a model which becomes Eq.~\eqref{eq:model} when projected onto the low-energy quasi-doublet.  We can directly relate $g_\text{eff}$ used in Eq.~\eqref{eq:model} to the out-of-plane $g$-factor, as $g_c\approx g_\text{eff}/10.9\approx 1.1$, which is close to the predicted Land\'e $g$-factor. The in-plane $g$-factor is an added free parameter, which needs to take the value $g_{ab}\approx2.4$ to reproduce the data.

Our interpretation, then, is that \ch{TmZn2GaO5} is a material lacking a temperature- or magnetic field-driven phase transition, yet it develops a ``hidden'' multipolar polarization at low-temperatures and has measurable and well-defined magnons.

\begin{figure}[htbp]
    \centering
    \includegraphics[width=9cm]{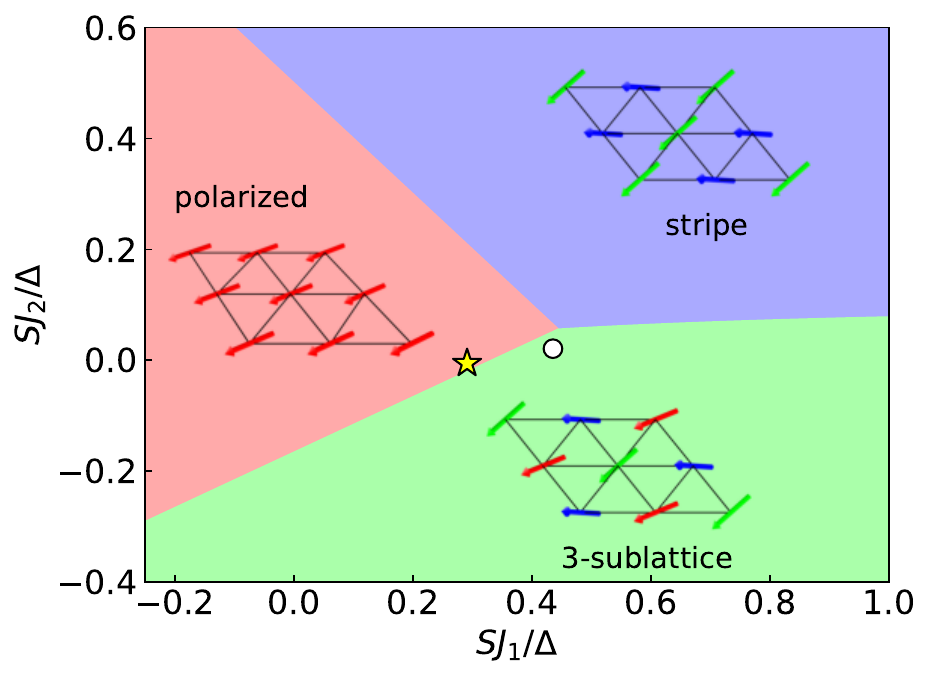}
    \caption{Phase diagram of classical orders on the triangular lattice: stripe (blue region), polarized (red region) and three-sublattice (green region). The star represents \ch{TmZn2GaO5}, and the dot represents the sister compound \ch{TmMgGaO4} for reference.}
    \label{fig:phasediagram}
\end{figure}

In conclusion, we have successfully synthesized, grown single crystals, and characterized \ch{TmZn2GaO5}, a novel rare-earth compound with a triangular lattice geometry, as a promising candidate for hosting an exotic magnetic ground state. Magnetic susceptibility, heat capacity, and neutron scattering measurements confirm the absence of long-range magnetic order down to 0.06 K, while magnetization reveals significant easy-axis anisotropy along the crystallographic $c$-axis. Inelastic neutron scattering identifies a low-energy gapped excitation emerging from the K point, aligning well with theoretical predictions. Interestingly, despite the lack of conventional temperature- or field-driven phase transitions, \ch{TmZn2GaO5} develops a ``hidden'' multipolar polarization at low temperatures, accompanied by measurable and well-defined magnons. Our findings place \ch{TmZn2GaO5} in a unique region of the frustrated quantum phase diagram, distinct from \ch{TmMgGaO4}, highlighting the crucial role of its structural features in shaping its magnetic behavior.  The absence of detectable site disorder further strengthens its potential as a platform for probing QSLs and other emergent quantum phases. These results underscore the significance of structural modifications in tailoring magnetic ground states and pave the way for future studies of frustrated magnetism and anisotropic interactions in quantum materials.

We thank Wenda Si and SiJie Xu for their contributions during the early stages of this project. The work performed at Duke University is supported by the U.S. Department of Energy, Office of Science, Office of Basic Energy Sciences, under Award Number DE-SC0023405. A portion of this research used resources at the Spallation Neutron Source and High Flux Isotope Reactor, DOE Office of Science User Facilities operated by the Oak Ridge National Laboratory. R.B. acknowledges the support provided by Fritz London Endowed Post-doctoral Research Fellowship at Duke University. T.C. is supported by a University of California Presidential Postdoctoral Fellowship and acknowledges support from the Gordon and Betty Moore Foundation through Grant No. GBMF8690 to UCSB. T.C. and L.B. were supported by the the NSF CMMT program under Grant No. DMR-2419871, and by the Simons Collaboration on Ultra-Quantum Matter, which is a grant from the Simons Foundation (651440).

\section{Author Contributions}
Research conceived by S.H.; M.E., R.B. and S.H. synthesized samples; M.E., R.B. and S.H. performed thermodynamics measurements; M.E., R.B., A.I.K, M.B.S, T.H. and S.H. conducted neutron scattering experiments; T.C. and L.B. provided theoretical interpretations; M.E., R.B., T.C., L.B. and S.H. wrote the manuscript with comments from all authors.

\bibliographystyle{apsrev4-2}
\bibliography{tm1215}

\end{document}


\title{Supporting Information for "Emergent Hidden Multipolar State in the Triangular Lattice Magnet \ch{TmZn2GaO5}"}

\author{Matthew Ennis}\thanks{equal contribution}\affiliation{Department of Physics, Duke University, Durham, NC, USA}
\author{Rabindranath Bag}\thanks{equal contribution}\affiliation{Department of Physics, Duke University, Durham, NC, USA}
\author{Tessa Cookmeyer}\thanks{equal contribution}\affiliation{Kavli Institute for Theoretical Physics, University of California, Santa Barbara, CA, USA}
\author{Matthew B. Stone}\affiliation{Neutron Scattering Division, Oak Ridge National Laboratory, Oak Ridge, TN, USA}
\author{Alexander I. Kolesnikov}\affiliation{Neutron Scattering Division, Oak Ridge National Laboratory, Oak Ridge, TN, USA}
\author{Tao Hong}\affiliation{Neutron Scattering Division, Oak Ridge National Laboratory, Oak Ridge, TN, USA}
\author{Leon Balents}\affiliation{Kavli Institute for Theoretical Physics, University of California, Santa Barbara, CA, USA}
\author{Sara Haravifard}
\email[email:]{sara.haravifard@duke.edu} \affiliation{Department of Physics, Duke University, Durham, NC, USA} \affiliation{Department of Mechanical Engineering and Materials Science, Duke University, Durham, NC, USA}
\affiliation{Department of Electrical and Computer Engineering, Duke University, Durham, NC, USA}


\maketitle

\section{Phase characterization and single crystal growth}
\label{Crystal Structure}

\begin{figure}
\centering
\includegraphics[width=12 cm]{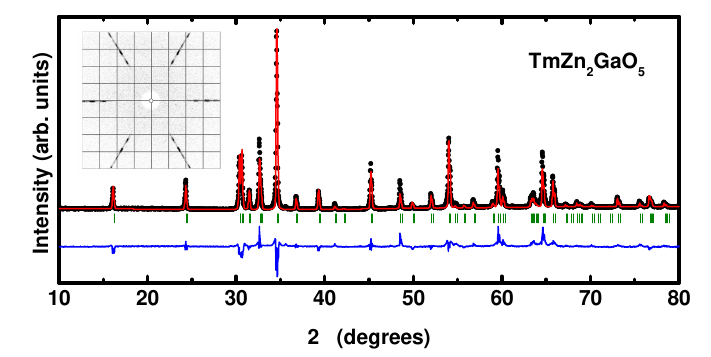}
\caption{Powder x-ray diffraction measurement of \ch{TmZn2GaO5} (black points) along with Rietveld fit to the $\mathrm{P}6_3/mmc$ space group (red curve) and the residual (blue curve). The inset shows a representative Laue pattern viewed along the (001) direction.}
\label{fig:PXRD}
\end{figure}

The powder samples of \ch{TmZn2GaO5} were synthesized via solid-state reactions using high purity precursors of \ch{Tm2O3}, \ch{ZnO}, and \ch{Ga2O3}. The starting materials were mixed in stoichiometric ratios, with an additional 5\% by weight of \ch{ZnO} added to minimize the residual \ch{Tm2O3} in the final product. The powder mixture was ground thoroughly in a mortar and pestle, pressed into a pellet, and sintered in air for 1100 $^\circ$C for 12 hours, followed immediately by sintering at 1350 $^\circ$C for 60 hours. The final synthesized powder is obtained as pure phase of \ch{TmZn2GaO5}. A representative powder x-ray diffraction pattern is shown in Fig. \ref{fig:PXRD} along with the Rietveld refinement confirming the pure phase of \ch{TmZn2GaO5} with $\mathrm{P}6_3/mmc$ space group.

To grow single crystal samples of \ch{TmZn2GaO5}, pure powder samples were pressed into cylindrical rods (feed and seed) using a hydrostatic press. These rods were sintered at 1200 $^\circ$C for 24 hours to strengthen them. Single crystal growth was performed in a four mirror optical floating zone furnace, with slow growth rates found to be essential for obtaining large single grain. The optimal growth parameters were moving the seed and feed rods at 1.05 mm/hr and 1 mm/hour respectively, with both rods rotating at between 10 and 15 rpm.

Due to porosity of the ceramic rods, liquid absorption into feed rod destabilized the melt during the slow growth process. To counteract this, a premelting step was carried out with with feed and seed speeds of 90 mm/hr and a rotation rate of 15 rpm. This produced a polycrystal that was much denser than the starting rod. This premelted rod was then used under slower growth conditions, significantly stabilizing the process and obtained high purity single crystals of \ch{TmZn2GaO5}. The grown crystals were oriented using Laue diffractometer and used for heat capacity and magnetic characterization. Additionally, four crystal pieces were co-aligned for inelastic neutron scattering.

\section{Additional magnetic data}
\begin{figure}
\centering
\includegraphics[width=15 cm]{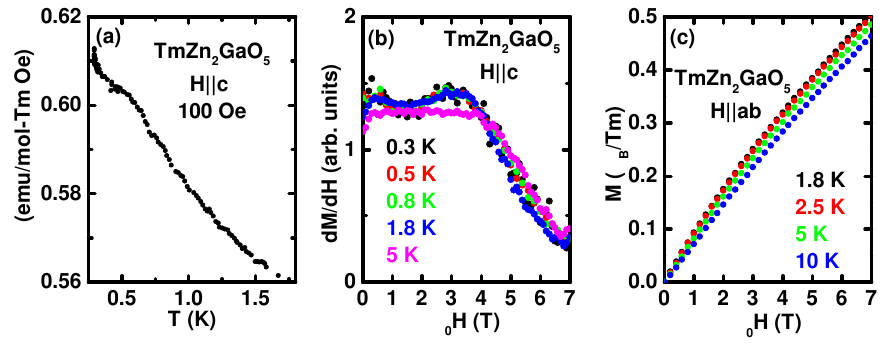}
\caption{(a) Magnetic susceptibility with 100 Oe field applied along $c$-direction down to 0.3 K (b) Field derivative of magnetization ($dM/dH$) for fields applied along $c$-axis, highlights oscillatory behavior at low fields (c) Detailed isothermal magnetization data for fields applied in-plane, showing almost linear behavior upto 7 T of magnetic field.}
\label{fig:mag_supplemental}
\end{figure}

In Fig. \ref{fig:mag_supplemental}, we show some additional magnetic data collected on \ch{TmZn2GaO5} single crystal sample. In the main text, Fig. 1(c) shows the magnetic susceptibility data for both orientations down to 1.8 K. Additionaly, we measured susceptibility for $H\parallel c$ down to 0.3 K as shown in Fig. \ref{fig:mag_supplemental}(a). Below 1.8 K, no  magnetic ordering is observed, and the susceptibility continues to rise down to 0.3 K. 

In Fig. 1(d) of the main text, the magnetization data exhibit small temperature dependence below 2.5 K, along with several slope changes before beginning to saturate above 4 T. More clearly see these slope changes in the magnetization, we have calculated the derivative $dM/dH$ for temperatures below 5 K as shown in Fig. \ref{fig:mag_supplemental}(b). Between 0 and 4 T, $dM/dH$ displays oscillatory behavior, with maxima at 0.4 T and 3 T, and a minimum at 1.4 T. At temperatures above 1.8 K, the maxima are suppressed, leading to flat behavior in the 5 K data. Above 4 T, the slope decreases monotonically for all temperatures, although the system still has not fully saturated by 7 T. 

In contrast, no similar features are observed in the magnetization data with field applied in-plane ($H \parallel ab$). Due to the significant anisotropy, overplotting both orientations obscure the details, therefore, the $H \parallel ab$ data is plotted separately in Fig. \ref{fig:mag_supplemental}(c). It shows minimal temperature dependence, similar to $H \parallel c$ data, but remains almost linear across the entire field range. We observed significant anisotropy in the magnetization data and unlike the $H \parallel c$ orientation, no slope changes are seen, and in plane in-plane saturation field is substantially higher than 7 T.
 
\section{Details of CEF measurements and fit}

\begin{figure}
\centering
\includegraphics[width=12 cm]{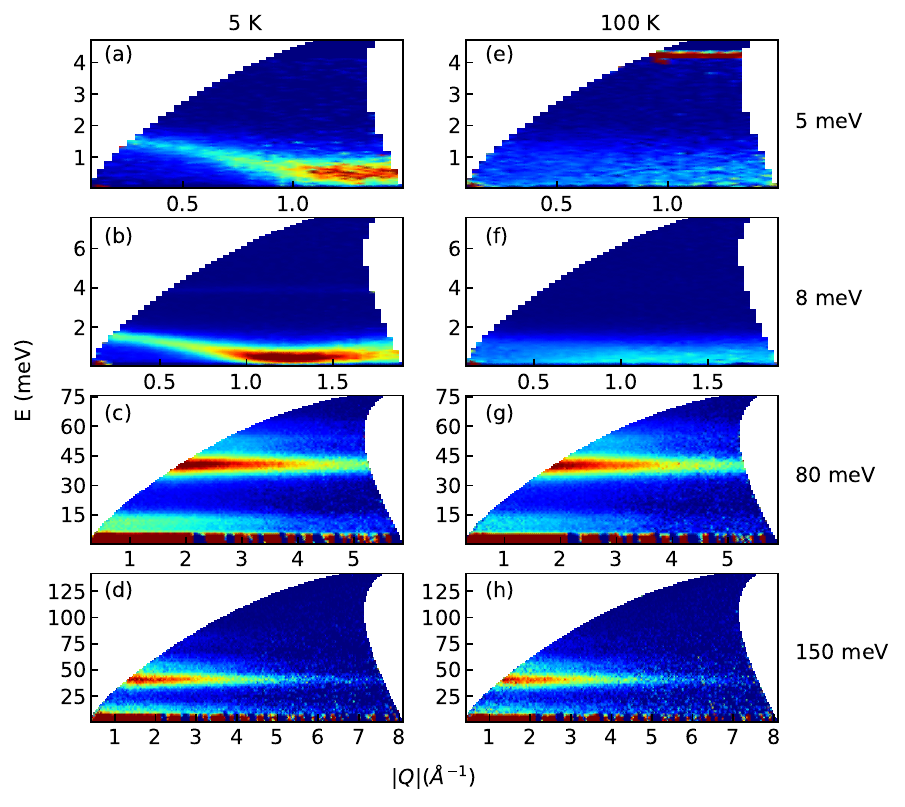}
\caption{Powder inelastic neutron scattering data of \ch{TmZn2GaO5} collected using the SEQUOIA spectrometer at ORNL. Panels (a-d) show measurements taken at 5 K, while panels (e-h) show measurements taken at 100 K. From the top to bottom, data corresponds to incident neutron energies of 5 meV, 8 meV, 80 meV, and 150 meV, respectively.}
\label{fig:SEQUOIA}
\end{figure}

To determine the crystal electric field (CEF) spectrum of \ch{TmZn2GaO5}, we conducted inelastic neutron scattering experiemnts on a powder sample with the SEQUOIA time-of-flight spectrometer at Oak Ridge National Laboratory \cite{SEQUOIA}. Measurements were performed at 5 K and 100 K with incident energy neutrons of 5 meV, 8 meV, 80 meV, and 150 meV. We used the high flux configuration for the 80 meV and 150 meV measurements, and the high resolution configuration for the 5 meV and 8 meV measurements. All of the powder neutron data are shown in Fig. \ref{fig:SEQUOIA}. 

As discussed in the main text, we observe a CEF-like excitation at 40 meV, with several weaker peaks at higher energies. No CEF excitations are detected below 40 meV or above 100 meV. For a \ch{Tm^{3+}} atom in the $D_{3d}$ environment, 10 energy levels (5 singlets and 4 doublets) are expected \cite{Li_PRX_2020}. However, only 3 distinct peaks are clearly observed, suggesting that these energy levels are closely spaced and blending into one another. Due to the difficulty of clearly identifying many close-lying states, we take the approach described by Bertin \textit{et al} for the rare-earth pyrochlores \cite{Bertin_2012}. 

\begin{table*}
\caption{Fitted CEF eigenvalues and eigenvectors of \ch{TmZn2GaO5}}
\begin{ruledtabular}
\begin{tabular}{c|ccccccccccccc}
E (meV) &$|-6\rangle$ & $|-5\rangle$ & $|-4\rangle$ & $|-3\rangle$ & $|-2\rangle$ & $|-1\rangle$ & $|0\rangle$ & $|1\rangle$ & $|2\rangle$ & $|3\rangle$ & $|4\rangle$ & $|5\rangle$ & $|6\rangle$ \tabularnewline
 \hline 
0.000 & -0.6475 & 0.0 & 0.0 & -0.2015 & 0.0 & 0.0 & -0.2832 & 0.0 & 0.0 & 0.2015 & 0.0 & 0.0 & -0.6475 \tabularnewline
0.150 & 0.6496 & 0.0 & 0.0 & 0.2794 & 0.0 & 0.0 & -0.0 & 0.0 & 0.0 & 0.2794 & 0.0 & 0.0 & -0.6496 \tabularnewline
40.857 & 0.0 & -0.5973 & 0.0 & 0.0 & -0.5949 & 0.0 & 0.0 & -0.5378 & 0.0 & 0.0 & 0.0077 & 0.0 & 0.0 \tabularnewline
40.857 & 0.0 & 0.0 & -0.0077 & 0.0 & 0.0 & -0.5378 & 0.0 & 0.0 & 0.5949 & 0.0 & 0.0 & -0.5973 & 0.0 \tabularnewline
51.397 & 0.0 & -0.1337 & 0.0 & 0.0 & -0.488 & 0.0 & 0.0 & 0.6807 & 0.0 & 0.0 & -0.5298 & 0.0 & 0.0 \tabularnewline
51.397 & 0.0 & 0.0 & 0.5298 & 0.0 & 0.0 & 0.6807 & 0.0 & 0.0 & 0.488 & 0.0 & 0.0 & -0.1337 & 0.0 \tabularnewline
60.108 & 0.2381 & 0.0 & 0.0 & -0.1131 & 0.0 & 0.0 & -0.9279 & 0.0 & 0.0 & 0.1131 & 0.0 & 0.0 & 0.2381 \tabularnewline
82.334 & -0.2794 & 0.0 & 0.0 & 0.6496 & 0.0 & 0.0 & 0.0 & 0.0 & 0.0 & 0.6496 & 0.0 & 0.0 & 0.2794 \tabularnewline
110.963 & 0.0 & 0.0 & 0.4404 & 0.0 & 0.0 & -0.4369 & 0.0 & 0.0 & 0.3267 & 0.0 & 0.0 & 0.7131 & 0.0 \tabularnewline
110.963 & 0.0 & -0.7131 & 0.0 & 0.0 & 0.3267 & 0.0 & 0.0 & 0.4369 & 0.0 & 0.0 & 0.4404 & 0.0 & 0.0 \tabularnewline
126.687 & 0.0 & 0.0 & -0.7248 & 0.0 & 0.0 & 0.2378 & 0.0 & 0.0 & 0.5488 & 0.0 & 0.0 & 0.3419 & 0.0 \tabularnewline
126.687 & 0.0 & -0.3419 & 0.0 & 0.0 & 0.5488 & 0.0 & 0.0 & -0.2378 & 0.0 & 0.0 & -0.7248 & 0.0 & 0.0 \tabularnewline
128.417 & 0.1549 & 0.0 & 0.0 & -0.6683 & 0.0 & 0.0 & 0.2424 & 0.0 & 0.0 & 0.6683 & 0.0 & 0.0 & 0.1549 \tabularnewline
\end{tabular}\end{ruledtabular}
\label{tab:cef_evals}
\end{table*}

As noted in the main text, the CEF Hamiltonian can be written as 

$$H_\mathrm{CEF} = \sum_{m,n} B_m^n O_m^n$$

\noindent where $O_m^n$ are the Stevens operators and $B_m^n$ are the CEF parameters. The CEF parameters can also be expressed as $B_m^n = A_m^n \langle r^n \rangle \theta_n$, where $\langle r^n \rangle$ is the expectation value of position and $\theta_n$ is a multiplicative factor, both of which have been tabulated for the different rare-earths \cite{Hutchings_1964, Edvardsson_1998}. Owing to the similarity between the CEF spectra of \ch{TmZn2GaO5} and \ch{TmMgGaO4}, we used to parameters reported in Ref. \cite{Dun_PRB_2021} as a more accurate starting point than a simple point charge fit. The CEF parameters were then refined to the 80 meV and 150 meV neutron data simultaneously to obtain the fit shown in the main text. The fit values and corresponding eigenvalues and eigenstates are shown in Tables \ref{tab:cef_evals} and \ref{tab:cef_params}.

\section{Elastic neutron scattering at high symmetry points}

As previously reported, the related compound \ch{TmMgGaO4} orders into a three-sublattice state at low-temperatures, and this is accompanied by increased scattering at the K points \cite{Shen2019, Dun_PRB_2021}. Conversely, if the material formed stripe order instead, this would be associated with magnetic Bragg peaks at the M points, as has been observed in the triangular lattice material \ch{AgNiO2} \cite{Wheeler_2009}. To further confirm the placement of \ch{TmZn2GaO5} in the phase diagram, we performed elastic rocking-curve scans at 0.06 K using the Cold Triple-Axis spectrometer (CTAX) at Oak Ridge National Laboratory. 

\begin{figure}
\centering
\includegraphics[width=12 cm]{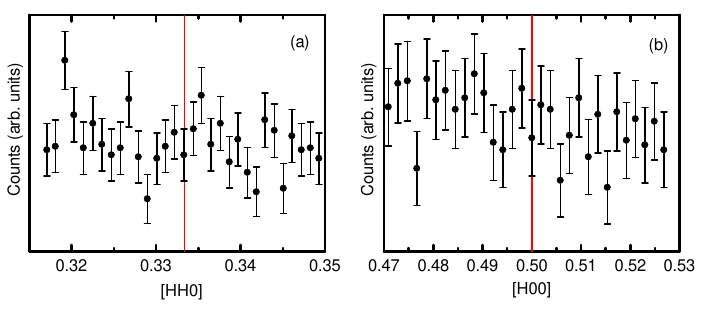}
\caption{Elastic rocking-curve scans performed at 0.06 K using CTAX, showing the data at (a) the K point and (b) the M point within the first Brillouin zone. The red lines indicate the position of the high-symmetry points. No peaks are observed at either point, ruling out the three-sublattice or stripe magnetic ordering.}
\label{fig:CTAX_elastic}
\end{figure}

Scans were conducted at the K and M points within the first Brillouin zone, as shown in Fig. \ref{fig:CTAX_elastic}. Due to the small size of the $a$ and $b$ lattice parameters (\~ 3.4 \AA) and limits for $Q$ coverage at CTAX, we were unable to perform measurements at the second Brillouin zone $\Gamma$ points to confirm if there was any increased scattering at the Bragg peaks at low temperatures. However, the absence of Bragg peaks at both K or M rules out the possibility of \ch{TmZn2GaO5} forming either the three-sublattice or stripe order, confirming our placement of it in the phase diagram.

\begin{table}
    \caption{CEF parameters for \ch{TmZn2GaO5}}
    \centering
    \begin{tabular}{|c|c|c|c|c|c|c|}
        \hline
        &  $B_2^0$ &  $B_4^0$ &  $B_4^3$ &  $B_6^0$ &  $B_6^3$ &  $B_6^6$ \\
        \hline
        $B_m^n$ (meV) & -0.344 & -0.00581 & -0.109 & -8.357$\times 10^{-6}$ & 3.555$\times 10^{-4}$ & -7.728$\times 10^{-4}$ \\
        \hline
    \end{tabular}
    \label{tab:cef_params}
\end{table}

\section{Theoretical analysis}

In the main text, we introduce the model Hamiltonian for spin-$1/2$s on a triangular lattice to describe the low-energy quasi-doublet:
\begin{equation}
    \mathcal{H} = \sum_{\langle ij \rangle} J_1S^z_iS^z_j + \sum_{\langle \langle ij \rangle \rangle} J_2S^z_iS^z_j + \sum_i \left(\Delta S^y_i - h_zS^z_i\right).
    \label{eq:model}
\end{equation} 
We will analyze this Hamiltonian at the classical, spin-wave, and mean-field level to interpret the experimental data. Importantly, we assume that the neutron scattering is only sensitive to the $S^z$ component of the spins as the $S^y$ and $S^x$ are time-reversal even \cite{Shen2019}.

As will be needed within some of the calculations, we choose the basis vectors of the triangular lattice to be $\vec a_1 = a(1,0)$ and $\vec a_2 = a(1/2,\sqrt{3}/2)$, where $a$ is the lattice constant. 

\subsection{Classical orders}

Because there is no energy gained from pointing the spins in the $x$ direction, it will choose to stay in the $y$-$z$ plane since it can always minimize energy by swapping any amount pointing in the $x$ direction for a larger component pointing in the $-y$ direction. Therefore, at the classical level, we just replace $S^z_i = S\cos(\theta_i)$ and $S^y_i = S\sin(\theta_i)$.

We consider all three and four spin unit cells to figure out what magnetic orders are possible. The energy $E=H/N$ per spin for the three spin unit cell is 
\begin{equation}
\begin{aligned}
    \frac{E_{3}}{S^2} &= \frac{J_1}{S^2} (S_1^z S_2^z + S_1^z S_3^z + S_2^z S_3^z) +  \frac{J_2}{S^2} [(S_1^z)^2 + (S_2^z)^2+(S_3^z)^2] - \frac{h_z}{3S^2} (S_1^z + S_2^z + S_3^z) + \frac{\Delta}{3S^2} (S_1^y + S_2^y + S_3^y) \\ 
    &=J_1(\cos(\theta_1) \cos(\theta_2) + \cos(\theta_1) \cos(\theta_3) + \cos(\theta_2) \cos(\theta_3) ) + J_2 ( \cos(\theta_1)^2 + \cos(\theta_2)^2 + \cos(\theta_3)^2) \\
    &-\frac{h_z}{3S} (\cos(\theta_1) + \cos(\theta_2) + \cos(\theta_3)) + \frac{\Delta}{3 S} (\sin(\theta_1) + \sin(\theta_2) + \sin(\theta_3))
\end{aligned}
\end{equation}
For four spin unit cells, there are two inequivalent boundary conditions that we consider. The first is a rhombus that is translated using $2\vec a_i$, where $\vec a_i$ the basis vectors of the triangular lattice:
\begin{equation}
\begin{aligned}
    \frac{E_{4,1}}{S^2} &= \frac{J_1 + J_2}{2 S^2} ( S_1^z S_2^z + S_1^z S_3^z + S_1^z S_4^z + S_2^z S_3^z + S_2^z S_4^z + S_3^z S_4^z) - \frac{h_z}{4 S^2} (S_1^z + S_2^z + S_3^z + S_4^z) + \frac{\Delta}{4 S^2} (S_1^y + S_2^y + S_3^y+ S_4^y) \\
    &= \frac{J_1 + J_2}{2} [\cos(\theta_1) \cos(\theta_2) + \cos(\theta_1) \cos(\theta_3) + \cos(\theta_1) \cos(\theta_4) + \cos(\theta_2) \cos(\theta_3) +\cos(\theta_2) \cos(\theta_4) + \cos(\theta_3) \cos(\theta_4) ] \\
    &- \frac{h_z}{4S} (\cos(\theta_1) + \cos(\theta_2) + \cos(\theta_3) + \cos(\theta_4)) + \frac{\Delta}{4S} (\sin(\theta_1) + \sin(\theta_2) + \sin(\theta_3) + \sin(\theta_4)).
\end{aligned}
\end{equation}
The other boundary condition has the rhombuses tile the plane with basis vectors $2\vec a_1$ and $2\vec a_2-\vec a_1$. We get
\begin{equation}
\begin{aligned}
    \frac{E_{4,2}}{S^2} &= \frac{J_1}{2 S^2} ( S_1^z S_2^z + S_1^z S_3^z + S_1^z S_4^z + S_2^z S_3^z + S_2^z S_4^z + S_3^z S_4^z) - \frac{h_z}{4 S^2} (S_1^z + S_2^z + S_3^z + S_4^z) + \frac{\Delta}{4 S^2} (S_1^y + S_2^y + S_3^y + S_4^y) \\
    &+\frac{J_2}{2} (S_1^z S_2^z + S_1^z S_4^z + S_2^z S_3^z + S_3^z S_4^z + \frac{1}{2}[(S_1^z)^2 + (S_2^z)^2 + (S_3^z)^2 + (S_4^z)^2]) \\
    &= \frac{J_1}{2} [\cos(\theta_1) \cos(\theta_2) + \cos(\theta_1) \cos(\theta_3) + \cos(\theta_1) \cos(\theta_4) + \cos(\theta_2) \cos(\theta_3) +\cos(\theta_2) \cos(\theta_4) + \cos(\theta_3) \cos(\theta_4) ] \\
    &- \frac{h_z}{4S} (\cos(\theta_1) + \cos(\theta_2) + \cos(\theta_3) + \cos(\theta_4)) + \frac{\Delta}{4S} (\sin(\theta_1) + \sin(\theta_2) + \sin(\theta_3) + \sin(\theta_4)) \\
    &+\frac{J_2}{2} \Big[\frac{\cos(\theta_1)^2+\cos(\theta_2)^2+\cos(\theta_3)^2 
    + \cos(\theta_4)^2}{2} \\ &\qquad\qquad+ \cos(\theta_1) \cos(\theta_2) + \cos(\theta_1) \cos(\theta_4) +\cos(\theta_3) \cos(\theta_2)+\cos(\theta_3) \cos(\theta_4)\Big]
\end{aligned}
\end{equation}

We can directly minimize these expressions with respect to the $\theta_i$'s. In the region we consider $J_1,J_2\gtrsim 0$, only three orders emerge: polarized, three-site, and stripe.

\textbf{(1) Polarized:} All the spins are polarized in the $-\hat y$ direction with $\theta_i = 3\pi/2$

\textbf{(2) Three-site:} The spins are primarily polarized but have a three-site unit cell with $\theta_1 = 3\pi/2$, $\theta_2 = 3\pi/2+\phi$ and $\theta_3 = 3\pi/2-\phi$. The angle $\phi$ varies depending on the parameters.

\textbf{(3) Stripe:} The spins are primarily polarized but have a two-site unit cell. The energy of this order is found using $E_{4,1}$ with $\theta_1=\theta_2 = 3\pi/2 -\phi$ and $\theta_3=\theta_4 = 3\pi/2+\phi$. 

We can evaluate the energy of these orders by analyticaly finding $\phi_\text{min}$ that minimizes the corresponding energy. We arrive at
\begin{equation}
\begin{aligned}
    E_{\text{three-site}}(\phi) &= -\frac{\Delta}{3}(2\cos(\phi)+1) -(J_1 -2 J_2)\sin(\phi)^2  \\
    E_{\text{stripe}}(\phi)&=-\Delta \cos(\phi) -(J_1+J_2)\sin(\phi)^2
\end{aligned}
\end{equation}
and
\begin{equation}
\begin{aligned}
    E_\text{polarized} &= -\frac{\Delta}{S}; \\
    E_{\text{three-site}}(\phi_\text{min}) &= 2 J_2 - J_1 - \frac{1}{9} \frac{\Delta}{S}\left(3+\frac{\Delta/S}{J_1 - 2J_2}\right) & \text{  if $\frac{\Delta/S}{3J_1 -6J_2}\le 1$} \\
    E_\text{stripe}(\phi_\text{min}) &=- \frac{(\Delta/S)^2 + 4(J_1 + J_2)^2}{4(J_1 + J_2)} & \text{ if $\frac{\Delta/S}{2J_1 + 2J_2}\le 1$}
\end{aligned}
\end{equation}
The conditions for $E_\text{three-site}(\phi_\text{min})$ and $E_\text{stripe}(\phi_\text{min})$ ensure that $\phi_\text{min}\ne 0$ corresponds to the minimum energy. By setting these expressions equal to each other, we can produce the classical phase diagram, as shown in the main text. 

\subsection{Spin-wave theory}

Since no magnetic Bragg peaks are observed, the ground-state configurations of the spins must be the polarized phase. We now evaluate the spin-wave excitations of this order. We rewrite all our spins as
\begin{equation}
    \label{eq:spinsub}
   \vec S = \begin{pmatrix} 1 & 0 &  0\\
   0 & \cos(\theta) & \sin(\theta) \\
   0 & -\sin(\theta)  & \cos(\theta) \end{pmatrix} \vec \Omega
\end{equation}
where $\vec S = (S_x, S_y, S_z)^T$ (and similarly for $\vec \Omega$, and $\Omega^z_i = S-b_i^\dagger b_i$, $\Omega^y_i =( b_i + b_i^\dagger)\sqrt{S/2}$ are expanded in terms of Holstein-Primakoff bosons. By inserting Eq.~\eqref{eq:spinsub} into Eq.~\eqref{eq:model}, we will gather terms with equal numbers of boson operators. We find that $H/N \approx H_0 S^2 + H_1 S^{3/2} + H_2 S + ...$ where
\begin{equation}
\begin{aligned}
    H_0 &=\frac{6J_1+6J_2}{2}  \cos^2(\theta) +\frac{\Delta}{S}  \sin(\theta) -\frac{h_z}{S} \cos(\theta) \\
    H_1 &= \frac{1}{\sqrt{2}N} \sum_i \left[\left(\frac{h_z}{S} \sin(\theta)+\frac{\Delta}{S} \cos(\theta)\right) -( 6J_1 + 6J_2)\sin(\theta) \cos(\theta)\right](b_i + b_i^\dagger)\\
    H_2 &= \frac{1}{N} \sum_i \frac{J_1}{2} \sum_\alpha \frac12\sin^2(\theta)(b_i+b_i^\dagger)(b_{i+\alpha} + b_{i+\alpha}^\dagger)  -  \cos^2(\theta) (n_i + n_{i+\alpha})\\
    &+\frac{J_2}{2} \sum_\beta \frac12\sin^2(\theta)(b_i+b_i^\dagger)(b_{i+\beta} + b_{i+\beta}^\dagger)  -  \cos^2(\theta) (n_i + n_{i+\beta}) \\
    &-\sum_i \left(\frac{\Delta}{S}\sin(\theta) - \frac{h_z}{S} \cos(\theta\right) b_i^\dagger b_i
\end{aligned}
\end{equation}
where  $\alpha \in \{ \vec a_1, \vec a_2, \vec a_2 - \vec a_1, -\vec a_1, - \vec a_2, \vec a_1 - \vec a_2\}$ are the nearest-neighbor vectors and $\vec a_1 = a(1,0)$ and $\vec a_2 = a(\frac{1}{2},\frac{\sqrt{3}}2)$, as above. Similarly, the next-nearest-neighbor vectors are $\beta \in \{\vec a_1 + \vec a_2, -\vec a_1 - \vec a_2, 2\vec a_1 - \vec a_2, \vec a_2 - 2 \vec a_1, 2\vec a_2 - \vec a_1, \vec a_1 - 2\vec a_2\}$. $\theta$ is set by the condition that $H_1=0$, which recovers the classical energy $H_0$. We then Fourier transform $H_2$ to get
\begin{equation}
\begin{aligned}
    H_2 &= \frac{1}{N} \sum_p A(p) b_p^\dagger b_p + B(p) b_{-p} b_p + B(p)^* b_{-p}^\dagger b_p^\dagger \\
    &= \frac{1}{2N} \sum_p \begin{pmatrix} b_p^\dagger & b_{-p}\end{pmatrix} \begin{pmatrix} A(p) & [B(p) + B(-p)]^* \\
    [B(p) + B(-p)] & A(-p)
    \end{pmatrix} \begin{pmatrix} b_p \\ b_{-p}^\dagger\end{pmatrix} - A(-p)\\ 
    A(p) &= -\frac{\Delta}{S} \sin(\theta) + \frac{h_z}{S}\cos(\theta) - J_1 \left(6\cos^2(\theta) -\frac12\sin^2(\theta) \sum_\alpha \cos(p\alpha) \right)-J_2 \left(6\cos^2(\theta) -\frac12\sin^2(\theta) \sum_\beta \cos(p\beta) \right) \\
    B(p) &= \frac{J_1}{4} \sin^2(\theta)\sum_\alpha e^{ip \alpha} + \frac{J_2}{4} \sin^2(\theta) \sum_\beta e^{ip \beta} = J_1 \frac12\sin^2(\theta) \sum_{\alpha'} \cos(p\alpha) + J_2\frac12 \sin^2(\theta) \sum_{\beta'} \cos(p \beta),
\end{aligned}
\end{equation}
where $\alpha' \in \{\vec a_1,\vec a_2, \vec a_2 - \vec a_1\}$. 

The dispersion is given by the eigenvalues of 
\begin{equation}
    M(p) = \begin{pmatrix} A(p) & [B(p) + B(-p)]^* \\
    -[B(p) + B(-p)] & -A(-p)
    \end{pmatrix} = \begin{pmatrix} A(p) & 2B(p) \\
    -2 B(p) & -A(p)
    \end{pmatrix}
\end{equation}
which has pairs of eigenvalues $\pm E(p)$. We just evaluate the dispersion at the high-symmetry points. In our conventions, these points are $\Gamma = (0,0)$, $M=a^{-1}(0,2\pi/\sqrt{3})$ and $K = a^{-1}(2\pi/3, 2\pi/\sqrt{3})$. The energy of the spin-wave mode is simply 
\begin{equation}
\begin{aligned}
    SE(\Gamma) &= \Delta \sqrt{6 \frac{J_1+J_2}{\Delta/S}  +1} \\
    SE(M) & =\Delta \sqrt{1-2 \frac{J_1+J_2}{\Delta/S} } \\
    SE(K) & = \Delta \sqrt{1-3 \frac{J_1-2J_2}{\Delta/S} }
\end{aligned}
\end{equation}
where $SE(\vec k)$ is the energy measured in the experiment (since the SWT Hamiltonian is $H_2 S$).
This shows that the dispersion becomes gapless at $M$ when the stripe phase is lower energy and gapless at $K$ when the three-sublattice order is lower in energy.

Using the single-crystal data, we estimate $[SE(\Gamma),SE(M), SE(K)] =[ 1.5, 0.6, 0.3]$, in meV, which gives $J_1 \approx 0.53$, $J_2 \approx -0.01$ and $\Delta \approx 0.91$. We then use \texttt{spinw} \cite{toth2015linear} to produce the single-crystal and powder spectra plots shown in the main text where just the $S^z$-$S^z$ correlator contributes to the signal. The spectra are convolved with a Gaussian of width $\Delta E = 0.4$ \text{meV}. 

\subsection{Mean-field theory}

In order to compare with the magnetization and specific-heat data, we use mean-field theory. We proceed by mean-field decoupling $S_i^z S_j^z = \langle S_i^z\rangle S_j^z + \langle S_j^z\rangle S_i^z - \langle S_i^z\rangle \langle S_j^z\rangle$ and assume $m = \langle S_i^z\rangle$. We have assumed that the parameters are chosen to be in the polarized phase so that all spins have the same expectation value. We get
\begin{equation}
    H_\text{MF} = - N3m^2J_\text{eff}+\sum_i \Delta S_i^y + (-h_z + 6 m J_\text{eff})S_i^z = -N3m^2 J_\text{eff} + \sum_i E S_i^{\hat n} 
\end{equation}
where $J_\text{eff}=J_1 + J_2$ and $6$ is the number of nearest and next-nearest neighbors and $N$ is the number of sites. In the last step, we moved to the eigenbasis of the single-site Hamiltonian with $E = \sqrt{\Delta^2 + (h_z - 6mJ_\text{eff})^2}$. We diagonalized the single-site Hamiltonian with the unitary
\begin{equation}
   \begin{pmatrix} \langle\frac{1}{2}| \\ \langle-\frac{1}{2}|  \end{pmatrix}=  \begin{pmatrix} \cos(\varphi) & - \sin(\varphi) \\
    \sin(\varphi) & \cos(\varphi) \end{pmatrix}\begin{pmatrix} \langle +|\\ \langle-|  \end{pmatrix}
\end{equation}
where $|\pm \frac{1}{2} \rangle$ is in the $z$ basis and we are using $S_i^y= \frac12\sigma^x$ (and therefore $S_i^x = -\frac{1}{2}\sigma^y$). The angle $\varphi$ is determined explicitly by the relations
\begin{equation}
    \sin(2\varphi) = \frac{\Delta}{E};\qquad  \cos(2\varphi) = \frac{-h_z + 6mJ_\text{eff}}{E}
\end{equation}

The mean-field Hamiltonian has the same form as the mean-field Ising model Hamiltonian, and we therefore find that the partition function is
\begin{equation}
    \mathcal Z_{MF} = e^{\beta N3 m^2J_\text{eff}}[2\cosh(\beta E/2)]^N.
\end{equation}

We can then evaluate
\begin{equation}
   \frac{1}{N}\sum_i  \langle S_i^{\hat n}\rangle = - \frac1{N\beta} \frac{1}{\mathcal Z_{MF}} \frac{\partial \mathcal Z_{MF}}{\partial E} = -\frac{1}{2} \tanh\left(\frac{\beta E}{2}\right).
\end{equation}
To evaluate $m$, we note
\begin{equation}
   \langle \sigma^z\rangle = \left\langle \left|\frac{1}{2}\right\rangle \left\langle \frac{1}{2}\right| - \left|-\frac{1}{2}\right\rangle \left\langle -\frac{1}{2}\right|\right\rangle = \cos(2\varphi) \langle \sigma^{\hat n}\rangle - \sin(2\varphi) \langle \sigma^{\hat n}_x \rangle 
\end{equation}
where $\sigma^{\hat n}_x = |+\rangle\langle -| + H.c.$ and $\langle \sigma^{\hat n}_x\rangle = 0$ for the thermal state. Therefore
\begin{equation}\label{eq:mft_sc}
    m = \langle S_i^z\rangle = \frac{1}{2}\frac{h_z -6m J_\text{eff} }{E} \tanh\left(\frac{\beta E}{2}\right).
\end{equation}

For given values of $h_z, J_\text{eff}$, we can find the value of $-1/2\le m \le 1/2$ that satisfies Eq.~\eqref{eq:mft_sc}. The magnetization 
(in untis of $\mu_B$ per Tm) is then simply given as $M = g_\text{eff} m$. We can additionally compute the heat capacity 
\begin{equation}
    C_V = -T \frac{\partial^2 F}{\partial T^2} 
\end{equation}
for $F=-\ln(\mathcal Z_{MF})/\beta$. When $h_z=0$, the self-consistent solution for all temperatures is simply $m=0$ and we recover the Schottky anomaly form. In general, we evaluate $F(T)$ and numerically compute $\partial^2 F/\partial T^2$.

\subsection{Computing in-plane magnetization data}

Although our effective spin-1/2 model for the low-energy quasidoublet, Eq.~\eqref{eq:model}, can quantitatively match the out-of-plane magnetization, magnetic heat capacity, and inelastic neutron scattering data, it cannot explain the in-plane magnetization data. In order to do so, we consider the simple model
\begin{equation}
    H' = \sum_{\langle ij\rangle} K_1 \pmb J_i \cdot \pmb J_j + \sum_{\langle\langle ij\rangle \rangle} K_2 \pmb J_i \cdot \pmb J_j +\sum_i H_{0,i} - g_c \mu_B J^z_i H^z - g_{ab} \mu_B(J^x_i H^x + J^y_i H^y)
\end{equation}
where $\pmb J_i = \pmb S_i + \pmb L_i$ is the total angular momentum operator for the Tm atoms and $H_{0,i}$ is the single-ion CEF Hamiltonian. If we let $U_0$ be the matrix defined in Table~\ref{tab:cef_params} and $\Lambda_0$ as the diagonal matrix containing the energy eigenvalues in Table~\ref{tab:cef_params} (with the low-energy doublet's splitting being replaced with $\Delta$ to match our effective model), then we can write $H_{0,i} = U_0^\dagger \Lambda_0 U_0$. In general, we do not expect that the multi-site interactions to take such a simple form, when written in terms of the $\pmb J_i$, but they have little effect on our particular calculation. 

If $P_0$ is the projection operator onto the low-energy quasi-doublet, we get that the effective Hamiltonian, $P_0 H' P_0$ is identical to Eq.~\eqref{eq:model} with $J_1 = K_1 f_z^2$, $J_2 = K_2 f_z^2$, and $g_\text{eff} = g_c f_z$ where $f_z \approx 10.8677...$. To arrive at that result, we note that 
\begin{equation}
    P_0 J^z P_0 = f_z S^x;\qquad  P_0 J^x P_0 = P_0 J^y P_0 = 0
\end{equation}
with $S^x$ is a spin-1/2 operator. Additionally, $P_0 H_{0,i}P_0 = \Delta S^z$, up to a constant. After relabeling the spin degrees of freedom, we arrive at Eq.~\eqref{eq:model}. Note that our choice of $g_\text{eff}=12$ corresponds to $g_c \approx 1.1$ very close to the Land\'e g-factor of $1.16$ for $J=6$ with $S=1$ and $L=5$, as expected for Tm$^{3+}$\cite{Shen2019}. 

We can once again use mean-field theory to analyze this system. In this case, we are interested in an in-plane magnetic field, which we take to be in the $x$ direction without loss of generality (and $H^y=H^z=0$). At the mean-field level, we have
\begin{equation}
    H'_\text{MF} = -3N K_\text{eff} (m^x)^2 + \sum_i H_{0,i} + (6K_\text{eff} m^x - g_{ab} \mu_B H^x) J_i^x 
\end{equation}
with $K_\text{eff}=K_1 + K_2$ which we can diagonalize numerically. We then compute $m^x=\langle J^x\rangle$ for a thermal density matrix $\rho = e^{-\beta H'_\text{MF}} / \text{Tr}[e^{-\beta H'_\text{MF}}]$ and look for the self-consistent solution. The magnetization is then given as $g_{ab} m^x$, similar to before. Performing the equivalent calculation for the field in the $z$ direction agrees with our calculation for the low-energy doublet with discrepancies at the percent level, and performing the calculation with the field in the $y$ direction gives equivalent results as the $x$ direction (with small discrepancies likely due to numerical inaccuracy in the entries of $U_0$).

We find that $g_{ab}\approx 2.4$ and $K_\text{eff} = (J_1+J_2)/f_z^2$ reproduces the experimental data, and changing $K_\text{eff}$ has a minimal effect on the result. This value of $g_{ab}$ is approximately a factor of 2 larger than the Land\'e g-factor value, and is therefore a reasonable order of magnitude. 

\bibliographystyle{apsrev4-2}
\bibliography{supplemental}